# Tunable wavefront control in the visible spectrum using low-loss chalcogenide phase change metasurfaces

*Parikshit Moitra\*, Yunzheng Wang, Xinan Liang, Li Lu, Alyssa Poh, Tobias W.W. Mass, Robert E. Simpson, Arseniy I. Kuznetsov\* and Ramon Paniagua-Dominguez\**

Parikshit Moitra, Xinan Liang, Tobias W.W. Mass, Arseniy I. Kuznetsov, Ramon Paniagua-Dominguez
Institute of Materials Research and Engineering, A*STAR (Agency for Science, Technology and Research), 138634, Singapore
E-mail: Moitra_Parikshit@imre.a-star.edu.sg; Arseniy_Kuznetsov@imre.a-star.edu.sg; Ramon_Paniagua@imre.a-star.edu.sg

Yunzheng Wang[⊥], Li Lu, Alyssa Poh, Robert E. Simpson
Singapore University of Technology and Design, 487372, Singapore
[⊥] Current Affiliation: Optics Research and Engineering, Shandong University, 266237, P. R. China

All-dielectric metasurfaces provide unique solutions for advanced wavefront manipulation of light with complete control of amplitude and phase at sub-wavelength scales. One limitation, however, for most of these devices is the lack of any post-fabrication tunability of their response. To break this limit, a promising approach is employing phase-change-materials (PCM), which provide a fast, low energy and non-volatile means to endow metasurfaces with a switching mechanism. In this regard, great advancements have been done in the mid infrared and near infrared spectrum using different chalcogenides. In the visible spectral range, however, very few devices have demonstrated full phase manipulation, high efficiencies, and reversible switching. Here, we experimentally demonstrate a tunable all-dielectric Huygens' metasurface made of antimony sulfide ($Sb_2S_3$) PCM, a low loss and high-index material in the visible spectral range with a large contrast (~0.5) between its amorphous and crystalline states. We show ~2π phase modulation with high associated transmittance and use it to create switchable beam steering and holographic display devices. These novel chalcogenide PCM metasurfaces have the potential to emerge as a platform for next generation spatial light modulators and to impact application areas such as tunable and adaptive flat optics, LiDAR, and many more.

## 1. Introduction

Recent advances in all-dielectric metasurfaces have witnessed successful demonstration of loss-less resonant manipulation of light at optical frequencies by utilizing the fundamental and higher order electric and magnetic Mie modes in high-index dielectric nanoresonators.[1–10] In this regard, complete wavefront control has been achieved by creating Huygens' metasurfaces where electric and magnetic dipole modes spectrally overlap for a particular aspect ratio of the resonators, leading to dipole moments of the same amplitude and phase that translate into high transmission and 2π optical phase control across the resonance wavelength.[3,6,11] A complete 2π optical phase control with high transmission paves the way toward many practical applications with high efficiencies, such as optical beam bending, holography, vortex-beam generation, meta-lenses etc.[2,4,12–15] A fundamental limitation of these designs, however, is that they are static in nature, i.e., post-fabrication dynamic tuning is not possible.

In recent years, there has been a drive toward overcoming this limitation by making metasurfaces tunable / active / reconfigurable.[16–18] Two popular approaches are, either active tuning of resonances by tuning the optical properties of the environment surrounding the metasurface or, by tuning the optical properties of the materials comprising the meta-atoms. In recent studies, liquid crystals[17,19–24], transparent conducting oxides such as indium-tin-oxide (ITO)[25–28], two-dimensional materials such as graphene and transition-metal dichalcogenides,[29–33] highly doped semiconductors,[34–36] phase transition materials such as vanadium dioxide ($VO_2$)[37] and chalcogenide phase change materials (PCMs),[38–43] have been studied as the switching media.

Chalcogenide PCMs, particularly GST, have been explored extensively to demonstrate dynamic resonance tuning[44] and wavefront control[45,46] for their multi-level,[47,48] nonvolatile, fast and reversible[49] switching from amorphous to crystalline states[50] with large contrast of refractive indices and long term stability.[49,51,52] Most demonstrations of tunable metasurfaces utilizing GST were carried out in the mid infrared frequencies because it exhibits low loss and a large refractive index contrast between its amorphous and crystalline states in this frequency band. In the telecommunications regime GST is typically used to modulate the amplitude of light because the bandgap of GST is programmable from 0.5 eV to 0.7 eV, which concomitantly causes changes in near-IR absorption.[43] However, because of its high absorption loss, GST is not suitable for use at higher frequencies. Attempts were made to lower the absorption of GST in the near-IR by doping GST with selenium (Se) to form a quaternary alloy Ge-Sb-Se-Te, which is commonly referred to as GSST.[53] An optimum alloy $Ge_2Sb_2Se_4Te_1$ exhibits very low absorption loss for wavelength longer than 1 μm.[51] However, lossless operation of PCM at visible frequencies is still elusive.

Recently, antimony trisulfide ($Sb_2S_3$) has been proposed as a suitable PCM in the visible spectrum for tunable optics.[54] Switching of $Sb_2S_3$ thin planar films or metasurfaces has been demonstrated with the application of mainly three different stimuli: first, application of heat, where the substrate is slowly heated to 300 ºC, which is above the amorphous-to-crystalline phase transition temperature, in a small chamber filled with argon (Ar) and held there for 1 hour;[55] second, electrical pulses ,[56,57] where voltage pulses have been applied across ITO micro-heaters to induce Joule heating facilitating the phase transition; third, laser pulses, where multi-level reversible switching with high cyclability is recorded using a continuous-wave laser for crystallization and a femtosecond laser for amorphization[58], as well as, millisecond and nanosecond pulses are used to induce full crystallization and amorphization, respectively.[59] In its amorphous form, $Sb_2S_3$ is a relatively lossless and high refractive index ($n_{am}$) material in the visible spectrum with $n_{am}$ varying from 3 to 3.5 across the spectrum, which makes it a suitable material to realize nanoresonators for all-dielectric metasurfaces. The crystalline phase has even higher refractive index ($n_{cr}$) compared to $n_{am}$, with an index contrast $\Delta n = n_{cr} - n_{am}$ ~0.5, making it a perfect platform for all-dielectric tunable nanophotonics. While tunable structural colors have recently been demonstrated by switching nanoresonators made of $Sb_2S_3$,[55,57] no attempts have been made so far to achieve free-space dynamic phase control with this material.

Here, we demonstrate a complete and tunable wavefront control in transmission using a $Sb_2S_3$ Huygens' metasurface. By carefully developing the nanoantenna design and fabrication process for this novel material, we first experimentally demonstrate fulfilling of Huygens' condition at visible frequencies. We further demonstrate non-volatile reversible switching of a beam steering device based on this concept. Incidentally, we also observe a reversible, large spectral tuning of bound-states-in-the-continuum (BIC) in the fabricated devices.

## 2. Nanofabrication of antimony trisulfide metasurfaces

The fabrication of the PCM-based metasurface starts with the RF-sputtering (AJA Orion 5) of an $Sb_2S_3$ thin film (160 nm) on an ITO (23 nm) coated glass substrate. This ITO thickness is optimal for avoiding charging during electron beam lithography without strongly affecting the metasurface resonances. The complex refractive indices of $Sb_2S_3$ in its amorphous and crystalline states (Figure 1a) are measured using spectroscopic ellipsometry (Woollam) and are used as the material properties in the simulations. The amorphous film becomes completely lossless for wavelengths longer than 600 nm, and the metasurface is designed to operate in this spectral range. The schematic of the metasurface is shown in **Figure 1**b, where *d*, *g* and *p* are the diameter of the nanoresonators, the gap between nanoresonators and the periodicity of the lattice (square in our case). The metasurface is patterned using electron beam lithography on a 300 nm thick polymethyl methacrylate (PMMA, 950K A5), which is a positive tone resist spin coated on the substrate. Circular patterns are created using the electron beam lithography, which give rise to hole patterns in the PMMA after development. Here, it is important to note that $Sb_2S_3$ is a soft material and in the absence of a protective coating, is sensitive to many commonly used developers, such as the salty developer (NaOH/ NaCl) or TMAH (Tetramethylammonium Hydroxide) for HSQ (Hydrogen Silsesquioxane) resist or o-xylene for ZEP (Zeon Electron beam Positive-tone) resist. Hence, PMMA is a suitable choice among the conventional resists because the developer (MIBK:IPA=1:3) does not react with the $Sb_2S_3$ film. The PMMA pattern is then transferred in to the PCM film using dry etching (Oxford Plasma Etch) of the material using chlorine ($Cl_2$) gas. The thickness of the PMMA is selected in such a way that it can withstand the etching process but without compromising the smallest feature size (*g*) of 80 nm. The scanning electron microscopy (SEM) image of metasurface with hole patterns in $Sb_2S_3$ film is shown in **Figure 1**c. For better impedance matching of the metasurface to the surroundings, a 200 nm thick PMMA layer is spin coated and soft baked on top of the fabricated metasurface and a 1 mm thick PDMS (poly-dimethylsiloxane) film is carefully mounted on top of the PMMA, so that the surrounding medium has an average refractive index of 1.45, which is close to that of $SiO_2$ (silicon dioxide).

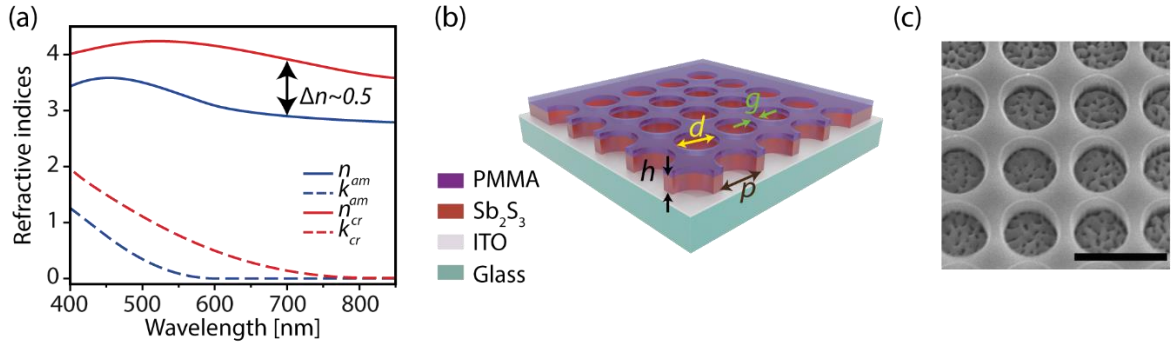

**Figure 1.** Metasurface design: Schematic and fabrication. a) Measured complex refractive indices for amorphous ($n_{am}$ & $k_{am}$) and crystallized ($n_{cr}$ & $k_{cr}$) film of $Sb_2S_3$. Large index contrast of $\Delta n \sim 0.5$ is apparent across the visible range. Amorphous $Sb_2S_3$ becomes lossless in the wavelength range longer than 600 nm, whereas the crystallized film becomes lossless beyond 800 nm. b) Schematic of hole metasurface. $d$, $g$, $h$ and $p=d+g$, are diameter of the nanoresonators, gap between the nanoresonators, height of the nanoresonators and periodicity of the metasurface, respectively, with $d=300$ nm, $g=80$ nm, $h=160$ nm, $p=380$ nm. c) Tilted (30º) SEM image of an array of hole nanoresonators made of amorphous $Sb_2S_3$. Scale bar is 500 nm.

## 3. Tunable Huygens' Metasurface: Design & Characterization

In this work, a metasurface design consisting of an array of nanoscale holes (Figure 1c) is studied. The reason for adopting this design over isolated nano resonators is two-fold: first, interconnected nanostructures allow easy propagation and growth of the crystal domains upon crystallization, second, the nanoholes still allow the Huygens' condition to be realized, as reported for other material platforms in different spectral bands.[60,61] Numerical simulations are performed using the finite difference time domain (FDTD, Lumerical) method to solve Maxwell's equations, where the hole diameters ($d$) are varied from 200 nm to 400 nm to achieve optimized Huygens' condition (Methods). **Figure 2**a,b show the simulated transmittance color maps with varying diameters of the nanoholes and wavelengths for amorphous and crystalline metasurfaces, respectively and **Figure 2**e,f show the phase colormaps of the same. The surrounding medium is considered uniform with a constant refractive index ($n = 1.45$), corresponding to the glass substrate and PMMA/PDMS superstrate. Huygens' condition, where both electric and magnetic dipole modes overlap and destructively interfere in the backward direction to exhibit high transmission, is designed to appear at 705 nm for the amorphous-phase metasurface with hole diameters ($d$) of 300 nm. The spectrum redshifts significantly, i.e., more than 100 nm, when the metasurface crystallizes. This effect is due to refractive index substantially increasing during the phase transition.

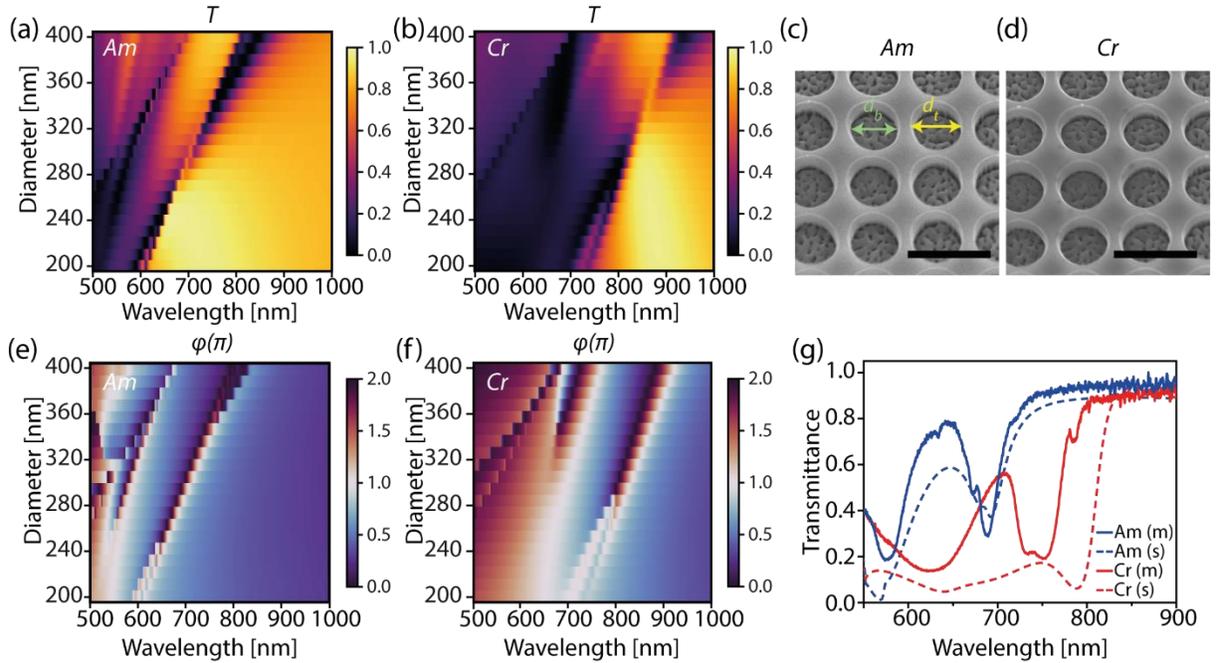

**Figure 2.** Tunable Huygens' Metasurface: Simulation and characterization. a) and b) Simulated transmittance color map as a function of diameter of nano holes and wavelength for amorphous (*Am*) and crystalline (*Cr*) metasurfaces, respectively. Huygens's condition is satisfied for amorphous metasurface with $d=300$ nm holes. Spectrum from the crystalline metasurface is significantly redshifted from that of amorphous one. Simulated phase maps for the same are given in e) and f). c) and d) Comparison of SEM images in the same location of the fabricated hole metasurfaces of amorphous-$Sb_2S_3$ post-fabrication and heat-treated crystallized-$Sb_2S_3$. No significant differences are observed in the nanostructures among the two different states. The nanoholes have tapering of 5⁰ from a straight side wall, which makes the bottom diameter ($d_b$) and top diameter ($d_t$) equal to 300 nm and 330 nm, respectively. The scale bars correspond to 500 nm. g) Comparison between the simulated (dashed line, abbreviated as '(s)' in the plot legend) and measured (solid line, abbreviated as '(m)' in the plot legend) spectra for amorphous (blue, abbreviated as 'Am' in the plot legend) and crystalline (red, abbreviated as 'Cr' in the plot legend) metasurfaces. The nanohole structure in simulations is considered the same as the fabricated nanoholes, i.e., bottom diameter ($d_b$) and top diameter ($d_t$) equal to 300 nm and 330 nm, respectively. There is a significant spectral shift of ~100 nm (simulation) and ~60 nm (measurement) between amorphous and crystalline spectra.

The transmission measurement is carried out using spectrally resolved back-focal-plane (BFP) imaging of the transmitted light through the metasurface (Methods), with incidence and collection objective of 50X magnifications with 0.55 NA (numerical aperture). Experimentally measured transmittance through the amorphous metasurface for near-normal incidence has a good agreement with the simulation, as shown in **Figure 2**g, for a metasurface with top diameter ($d_t$) and bottom diameter ($d_b$) of the nanoholes equal to 330 nm and 300 nm (Figure 2c,d), respectively, which is considered in the simulation as well. This small difference in $d_t$ and $d_b$ arises from the etching process with the side walls having a small angular deviation of 5⁰ from normal. The Huygens' condition in the measured spectrum appears near 680 nm, because of the said difference in the morphology of the nanoholes. To crystallize the metasurface, the sample is heated at 5⁰ C/min in a small furnace in a 5 sccm (standard cubic centimeters per minute) of argon (Ar) atmosphere. The sample is fully crystalized at 320⁰ C (Methods). Both the phases viz. amorphous and crystallized, are corroborated using Raman characterization (Figure S1, Supporting Information. The measured transmission from the crystallized metasurface (Figure 2g) shows a large (nearly 60 nm) spectral redshift that has a close resemblance to the simulated spectrum. **Figure 2**c,d compare the SEM images in the same location of one of the metasurfaces before and after crystallization. There is some roughness visible in the background, which may have generated from the non-reactive bombardment of the plasma on the ITO film during pattern transfer using reactive ion etching. This roughness, however, does not have any untoward effect on the resonances. The SEM images confirm that the amorphous and crystallized metasurfaces have no significant differences in their morphology. The large shift in the spectra is solely caused by the refractive index contrast between the amorphous and crystallized states.

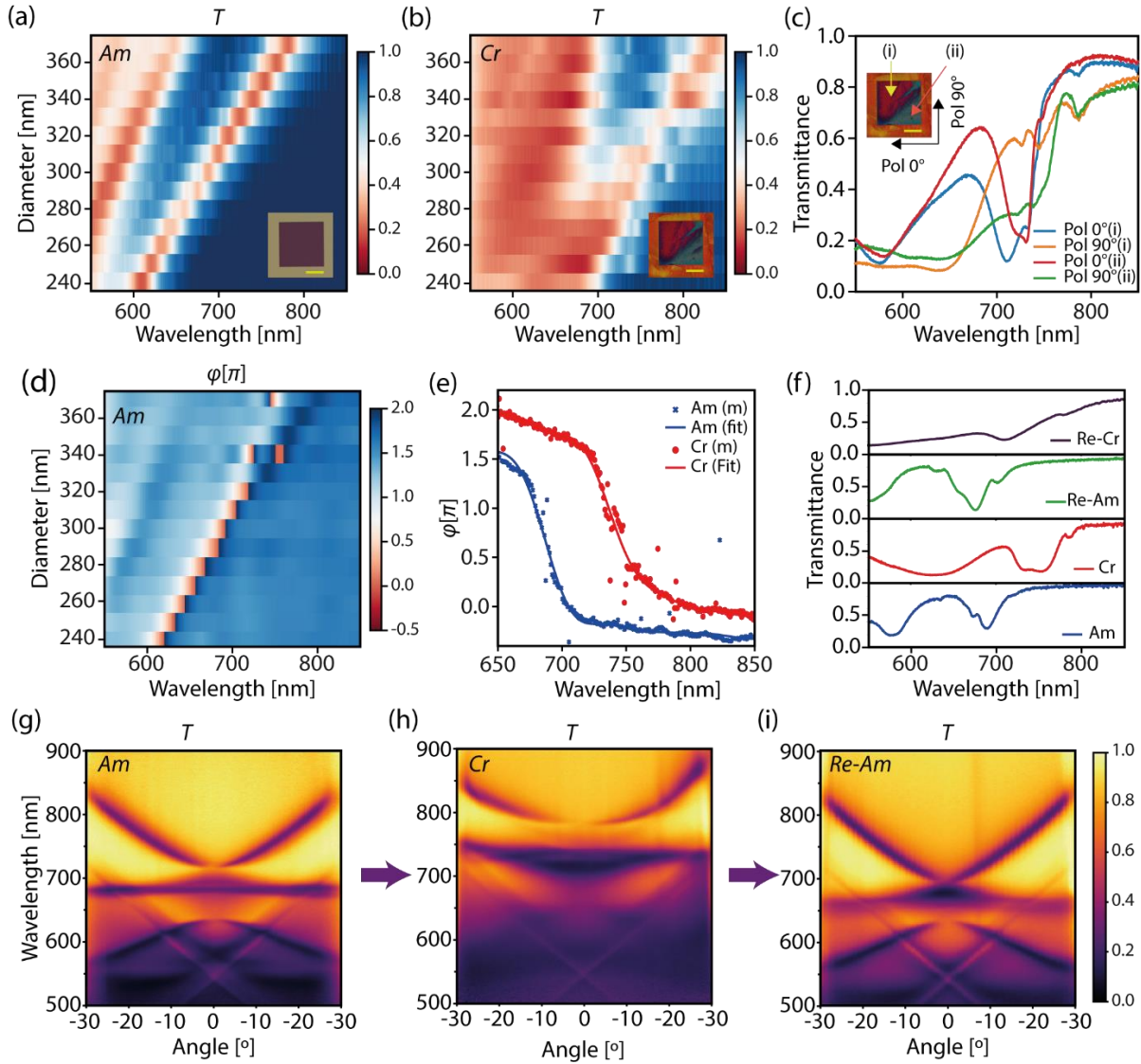

**Figure 3**. Tunable Huygens' metasurface and BIC. Measured transmission (*T*) maps for a) amorphous and b) crystallized $Sb_2S_3$ metasurfaces, as a function of bottom hole diameter $d_b$, varying from 240 nm to 370 nm with 10 nm increment, and wavelength. Optical images (transmission) of amorphous and crystallized arrays (50 μm × 50 μm) are shown in the inset. Crystallized array shows distinct crystallographic grains. c) Transmission comparison from different crystallographic grains ((i) and (ii)), from the same array ($d_b$=300nm), for two mutually

perpendicular incident polarizations (Pol 0º and Pol 90º) of light. d) Measured transmission phase map ($\varphi(\pi)$) of amorphous-$Sb_2S_3$ metasurfaces, as a function of bottom hole diameter $d_b$, varying from 240 nm to 370 nm with 10 nm increment, and wavelength. e) Comparison of measured phases from amorphous ('Am (m)', blue dots) and crystallized ('Cr (m)', red dots) metasurfaces with $d_b$=300 nm. Solid lines represent fitted phase points. f) Comparison of transmission spectra for amorphous ('Am'), crystallized ('Cr'), re-amorphized with laser ('Re-Am') and re-crystallized with laser ('Re-Cr') metasurfaces with $d_b$=300 nm. g), h), i) Demonstration of tunable BIC with amorphous, crystallized and re-amorphized metasurfaces.

The measured transmittance spectral map of metasurfaces with $d_b$ varying from 240 nm to 370 nm, while $d_t$=$d_b$+30 nm, are in **Figures 3**a and b for the amorphous and crystallized states, respectively. As expected, the resonance red shifts with increasing hole diameters in the amorphous metasurface, with the Huygens' condition prevailing at $d_b$=300 nm. The resonance shift with respect to the variations of hole diameters is also obvious from the transmittance map in the crystallized metasurface. The crystallized phase of $Sb_2S_3$ has large birefringent grains,[62] and consequently the overall transmittance is a weighted average of the transmission through these grains. These grains cause a distribution of resonance shifts. The transmission optical microscope images of the Huygens' metasurface before and after crystallization are shown in the insets of **Figure 3**a,b. The inset in **Figure 3**b clearly shows the different crystal domains in the array, with two main large grains observed with different transmissive colors. **Figure 3**c shows the effect of polarization on the transmittance through these different crystallized grains. Two separate transmission measurements are carried out on each grain ((i) and (ii)) for two mutually perpendicular incident polarization states (Pol 0º and Pol 90º). As can be seen, all the four transmission spectra are unique, with the same polarization states showing some resemblance, which might be an indication of close, albeit dissimilar, crystal axis orientations. These measurements corroborate the birefringent nature of crystalline $Sb_2S_3$ and highlight the need for further research to either mitigate this effect or controllably make use of it.

The measured optical phase of the $Sb_2S_3$ Huygens' metasurfaces shifts by nearly $2\pi$ radians during crystallization. The transmission phase measurement was carried out using a Mach-Zehnder interferometer (Methods). The broadband measured phase, $\varphi(\pi)$, for amorphous metasurfaces with varying diameter ($d_b$) shows large phase shifts across the resonant wavelength (Figure 3d). The measured phase shift for amorphous and crystallized Huygens' metasurfaces

($d_b$=300 nm) is covering nearly $2\pi$ phase around the respective resonance wavelengths of 680 nm and 740 nm (Figure 3e). This full $2\pi$ phase control with high transmittance makes this system a good candidate for beam bending and holographic display devices. Note, firstly, the relative phase shift among the two metasurfaces viz. amorphous and crystallized at 680 nm is >$1.5\pi$, opening the possibility for fully reconfigurable devices if intermediate crystallization states would be accessible; secondly, for the range of wavelengths over which we can obtain close to $2\pi$ phase shift in the amorphous case, the phase in the crystalline state remains almost constant, allowing to switch off (on) any inscribed functionality in the amorphous state by crystallizing (amorphizing) the device, as will be shown next. Also note, in a similar fashion to the amplitude, the phase shift also depends on the crystal grain orientation. Details are given in Supporting Information (Figure S2), showing that there are slight differences between the phase shifts among the different grains.

The $Sb_2S_3$ metasurfaces can also be reversibly switched between their amorphous and crystalline states. The crystalline $Sb_2S_3$ metasurface film is re-amorphized and then re-crystallized using laser pulses (Methods). **Figure 3**f compares the spectra for two cycles of switching. The re-amorphized metasurface spectrum has a significant blue-shift from that of the crystallized state and it closely follows the spectrum of the as-deposited amorphous state. However, there is a small blueshift of the resonance in the re-amorphized metasurface compared to as-deposited amorphous state. This small difference is likely due to some sulfur loss during the melt-quenching process.[55] Note, before the switching processes (heating by laser pulses or in a microscope furnace), the PMMA and PDMS were removed and the metasurface was encapsulated with a 100 nm $SiO_2$ layer, which was deposited by plasma enhanced chemical vapor deposition (PECVD). **Figure 3**f also shows that, upon laser re-crystallization, the metasurface spectrum significantly redshifts again from that of the re-amorphized state.

The full mode dispersion of the metasurface at the Huygens' condition can be spectrally tuned upon crystallization and then fully recovered upon re-amorphization. the angle-resolved transmission spectra of the metasurface array with $d_b$=300 nm were studied in different structural phase states (Figure 3g,h,i). One can see that the full mode dispersion spectrum can be reversibly tuned by the crystallized-amorphous phase transition. We also see the tuning of the bound-state-in-the-continuum (BIC) condition. Indeed, the (s-polarized) angle-resolved spectra show a resonance caused by vertical magnetic dipole at oblique incidence, which undergoes a spectral narrowing and eventually disappears at 0° at 710 nm. This is a signature of a BIC forming. By switching from the amorphous to crystalline and then to the re-amorphized state, the BIC is shifted from 710 nm to 780 nm and back to 690 nm.

## 4. Tunable Beam Steering

To demonstrate the power of the phase-change Huygens' metasurface for dynamic manipulation of wavefronts, a gradient metasurface is fabricated to realize switchable beam steering into the transmission +1 diffraction order. The device was created by periodically repeating a supercell consisting of 8 nanoholes, with their diameters varying from 260 nm to 320 nm, and with the periodicity and height fixed at $p$=380 nm and $h$=160 nm, respectively (see Figure 4a,b for SEM images). The diameters of the holes are selected so that when the metasurface is in the amorphous phase, there is a $\pi/4$ phase difference between adjacent cells at the operating wavelength (~705 nm, see Methods and Figure S3). This means that the supercell accumulates a total $2\pi$ phase difference linearly across it. Upon crystallization, no significant phase gradient is expected at 705 nm, and thus we do not expect to see any significant diffraction from the crystallized metasurface.

The beam steering experiment is carried out using spectrally resolved back-focal-plane imaging of the light transmitted through the metasurface with close-to-normal incidence (Methods). A 50× objective lens with NA of 0.55 is used to collect the diffracted light. Maps of the diffraction angle, wavelength and normalized transmission intensity matrix through the metasurfaces for the amorphous (Figure 4c), crystallized (Figure 4d) and re-amorphized (Figure 4e) states clearly show beam bending in the amorphous phases and no beam bending in the crystalline phase. The normalized transmission into the three main diffraction orders (0,−1,+1) is shown in **Figure 4**f,g,h. The higher diffraction orders are significantly weaker and can be neglected. When the wavelength of the incident beam approaches the design beam bending wavelength of ~680 nm, the 0 and all other diffraction orders except +1 are strongly suppressed. The +1 order is, by contrast, enhanced, which corresponds to the transmitted beam deflection to a steering angle of approximately 14°. From **Figure 4**f, one observes that the transmission efficiency of the amorphous metasurface reaches a maximum value of around 17% at 680 nm, i.e., 17% of incident light is transformed into the desired (+1) order. Tunable beam bending is achieved by switching the metasurface from the amorphous to the crystalline state. As explain above, in this case no significant phase gradient is expected at 680 nm, which should lead to simple, direct transmission of light. Indeed, the beam bending efficiency is reduced to almost zero for the crystallized state of the metasurface, with most of the light channeled to the 0 order and negligible light diffracting into both ±1 orders. Upon laser re-amorphization of $Sb_2S_3$, the beam bending efficiency again increases considerably to almost 10%. As a last experimental demonstration of switchable phase control, we show in Supporting Information (Figure S3), a hologram whose efficiency can be controlled by crystallization-amorphization of the metasurface.

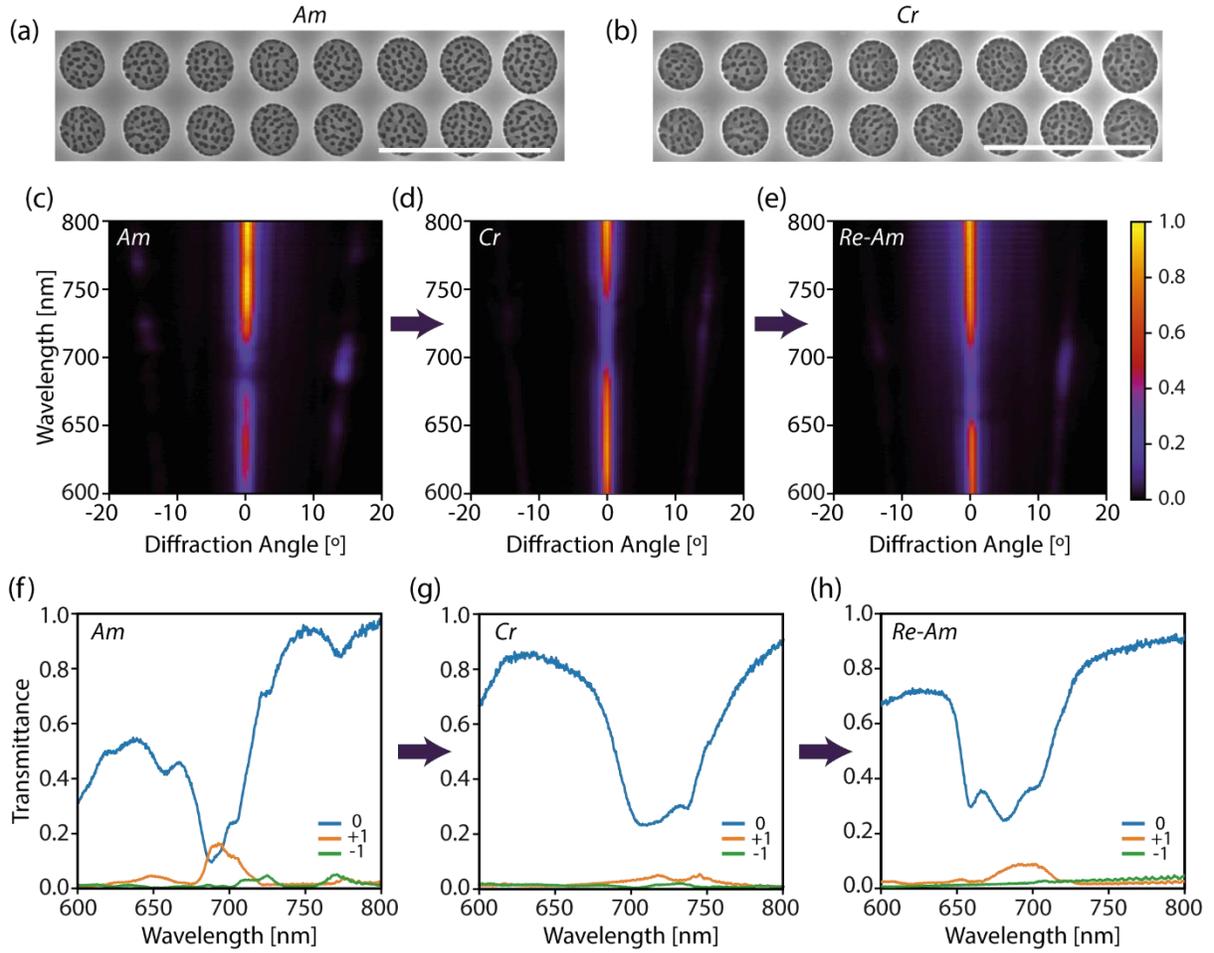

**Figure 4.** Tunable beam bending experiments. SEM images of the 8-elements supercell for a) amorphous (*Am*) and b) crystallized (*Cr*) $Sb_2S_3$ beam bending metasurfaces. Both images are taken from the same location in the array. The scale bars correspond to 1 µm. c) - e) Angle resolved normalized transmission spectra for amorphous, crystallized and laser re-amorphized metasurfaces. f)-h) show the normalized transmission power for each diffraction mode viz. 0, +1 and -1 order for amorphous, crystallized and laser re-amorphized metasurfaces.

## 5. Conclusion

Low loss and high refractive index of $Sb_2S_3$, together with its large index contrast (~0.5) between its amorphous and crystalline phases, makes it an ideal material to realize tunable all-dielectric metasurfaces operating in the visible spectrum. In this work, first, we developed the process to create high resolution nanoscale patterns using conventional nanofabrication techniques, bolstering the usefulness of this material. Pinning on this, we demonstrated Huygens' metasurfaces in which the nanoantennas are made of $Sb_2S_3$ and used them to realize tunable wavefront manipulation in the visible spectrum utilizing non-volatile and reversible switching between the amorphous and crystalline states. The advantages derived from the design, with the PCM comprising the metasurface and not just acting as a thin film surrounding the static resonators, are the large resonance shift of ~60 nm and transmission phase shift of more than $1.5\pi$ between the amorphous and crystallized states. This large change in optical phase directly enables phase-only manipulation of transmitted light with high efficiency. We used optical phase control to realize reversible switchable beam bending, holograms and even BICs. This demonstration paves the way to realizing high resolution spatial light modulators for AR / VR applications, tunable flat optics, LiDAR, and dynamic optical holography devices. Whilst undoubtedly attractive, the complete development of this approach will require further studies to overcome the appearance of large crystallographic grains upon crystallization, which may make the optical response non-optimal, and truly multi-level reversible switching, which until now has only been seen in continuous thin $Sb_2S_3$ thin films. We believe that, in this regard, our work is an important step toward realizing the next generation high resolution tunable optical devices in the visible spectrum.

## Methods

**Simulation methodology**

The metasurface is designed using full wave numerical simulations based on Finite Difference Time Domain method (FDTD, Lumerical Solutions). Periodic Boundary Conditions (PBC) are applied for the metasurface to be infinitely periodic in the in-plane directions. To achieve Huygens' condition the resonator diameters are swept in a wide range from 200 – 400 nm. The transmittance and phase are extracted and plotted using Python to achieve the transmission amplitude and phase color maps.

A gradient metasurface, to show beam bending, is designed with a supercell consisting of 8 nanoscale holes so that there is a $\pi/4$ phase difference between adjacent cells and the supercell accumulates $2\pi$ phase difference across it. The set of diameters comprising the supercell is 265, 275, 285, 290, 295, 300, 310 and 325 nm. The periodicity is fixed at 380 nm. See Supporting Information (Figure S3) for more details.

**Back focal plane imaging for transmission measurement**

Broadband white light from halogen light with *s* and *p*-polarizations, respectively, is incident on the metasurface using a Nikon, 50× 0.55 NA objective lens and the transmitted light through the metasurface is collected using another objective lens with the same specifications. The back focal plane (BFP) of the collection objective lens is imaged onto the 100μm entrance slit of a spectrograph (Andor Kymera 328i). Perpendicular to the slit, the light is dispersed by a blazed grating so that the back focal plane can be spectrally resolved in one direction. The image is collected by a Newton EMCCD-971 sensor array. The angle-resolved spectra are plotted by mapping the angle of incidence of light with the pixels in the sensor array. The line-spectra with normal incidence are shown by extracting the transmittance from the 0⁰ angle of incidence. A window with same size (50 μm × 50 μm) as the metasurface arrays is created in the substrate by completely etching away the film and is used as a reference for transmission spectrum normalization. A collimated light with normal incidence on the metasurface is used for the beam bending experiment.

**Phase measurements**

The phase shift $\varphi$ induced by the PCM metasurface is measured with a Mach–Zehnder interferometer and retrieved with a five-step temporal phase-shift algorithm.[63] A supercontinuum source (SuperK EXTREME, NKT Photonics) equipped with multi-wavelength filter (SuperK SELECT, NKT Photonic) is used as the laser source. The laser beam is split in to two beams viz. signal beam and reference beam, using a beam splitter. After passing through the sample, the signal beam carries the information of the metasurface and interfere with the reference beam resulting in forming of fringes, which are then imaged with a camera (Thorlabs, Quantalux 2.1 MP Monochrome sCMOS Camera). In the light path of the reference beam, a mirror, mounted on a piezo stage is used to achieve five-step phase shifts (0, 0.5π, 1.0π, 1.5π and 2.0π). The intensities i.e., $I_1$, $I_2$, $I_3$, $I_4$ and $I_5$ of the corresponding fringes are used to calculate the wrapped phase map according to the following equation:

$$\varphi = arctan \left[ \frac{2(I_2 - I_4)}{2I_3 - I_5 - I_1} \right]$$

After the wrapped phase is obtained, it is then processed with a sin-cos filter and unwrapped to obtain the final phase map of the PCM metasurface. A local reference i.e., a window with the same size as the arrays created by etching the $Sb_2S_3$ film, is used as the reference to calculate the relative phase-shift.

**Switching by heating and laser pulses**

The hole metasurfaces are first fully crystallized from (as-deposited) amorphous state by slowly heating (5 ⁰C/min) the substrate up to 320⁰ C in a Linkam furnace with controlled Ar environment. Then, the sample is installed on to a high-precision motorized xy-stage in a home-made triple-wavelength testing system to conduct subsequent laser-induced amorphization and crystallization cycles. The laser parameters for amorphization and crystallization are: λ = 660 nm, P = 28.6 mW, τ = 50 ns, f = 20 Hz, v = 10 μm/s and λ = 405 nm, and P = 24.1 mW, τ = 5 μs, f = 20 kHz, v = 20 μm/s, respectively, where λ is wavelength, P is laser power, τ is pulse width, f is pulse frequency, v is stage moving velocity. In both the cases the beam waist is around 0.8 μm.

**Acknowledgements**

This work was supported in part by the AME Programmatic Grant, Singapore, under Grant A18A7b0058; in part by the IET A F Harvey Engineering Research Prize 2016; and in part by the National Research Foundation of Singapore under Grant NRF-NRFI2017-01.

**Conflict of Interest**

The authors declare no conflict of interest.

**Data Availability Statement**

The data that support the findings of this study are available on request from the corresponding author. The data are not publicly available due to privacy or ethical restrictions

**Table of Contents**

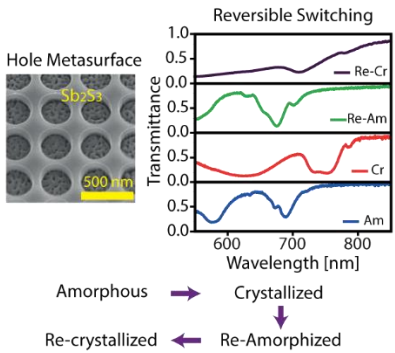

This work establishes antimony trisulfide ($Sb_2S_3$) as a good low-loss high-refractive index dielectric phase-change material to form all-dielectric metasurfaces for tunable wavefront control in the visible spectrum. The metasurfaces experimentally demonstrate reversibly tunable optical responses with large spectral shift and close to $2\pi$ phase modulation, along with tunable beam steering and bound state in the continuum.

# Supporting Information

**Tunable wavefront control in the visible spectrum using low-loss chalcogenide phase change metasurfaces**

*Parikshit Moitra\*, Yunzheng Wang, Xinan Liang, Li Lu, Alyssa Poh, Tobias W.W. Mass, Robert E. Simpson, Arseniy I. Kuznetsov\* and Ramon Paniagua-Dominguez\**

**Raman characterization**

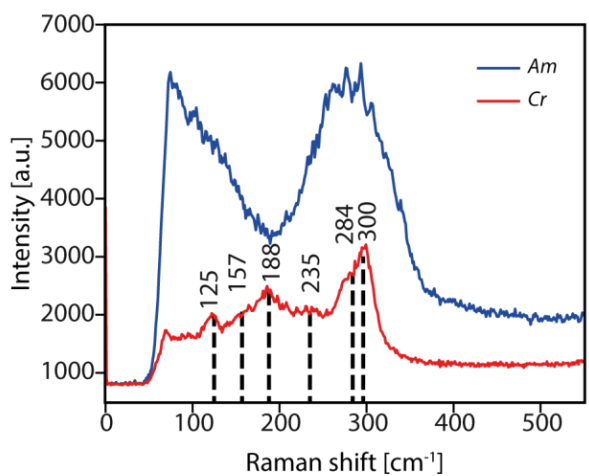

**Figure S1.** Raman spectra for amorphous and crystallized (crystallized by annealing) films of $Sb_2S_3$ demonstrating successful conversion from amorphous to crystallized state upon heat treatment.

Raman spectra for both the phases are presented in **Figure S1**. A linearly polarized 532 nm CW laser, coupled to a WITec Alpha300R Raman Imaging Microscope is used to record Raman signals from both amorphous and crystallized (annealing) $Sb_2S_3$ films. The amorphous film shows a Raman scattering characteristic with two broadband features at Raman shifts of 130 cm$^{-1}$ and 300 cm$^{-1}$, corresponding to Sb – S and S=S vibrational modes. Multiple peaks with more narrow linewidths emerge after annealing the $Sb_2S_3$ thin film, indicating a successful crystallization. While the vibrational mode around 188 cm$^{-1}$ corresponds to $Sb_2O_3$, arising from surface oxidization, the other vibrational modes indicated in **Figure S1** can be attributed to vibrational modes of different symmetry groups of crystalline $Sb_2S_3$.[1,2]

**Effect of crystal grains on optical phase of metasurfaces**

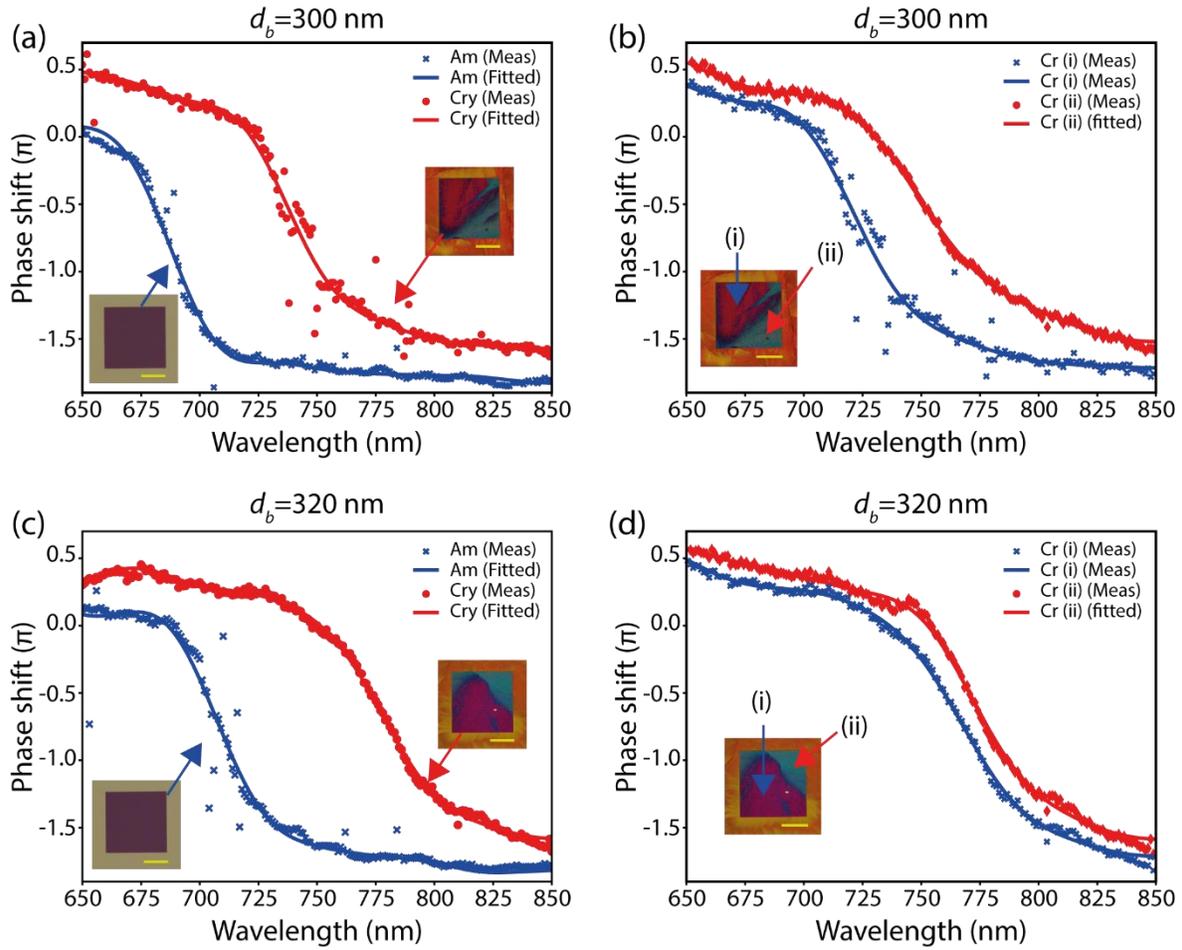

**Figure S2.** a) Phase shift ($\varphi(\pi)$) of amorphous and crystalline metasurface with $d_b$=300nm. b) Difference in phase shift between two crystallographic grains (i) and (ii) for $d_b$=300nm. c) Phase shift ($\varphi(\pi)$) of amorphous and crystalline metasurface with $d_b$=320nm. d) Difference in phase shift between two crystallographic grains (i) and (ii) for $d_b$=320nm.

**Tunable wavefront control**

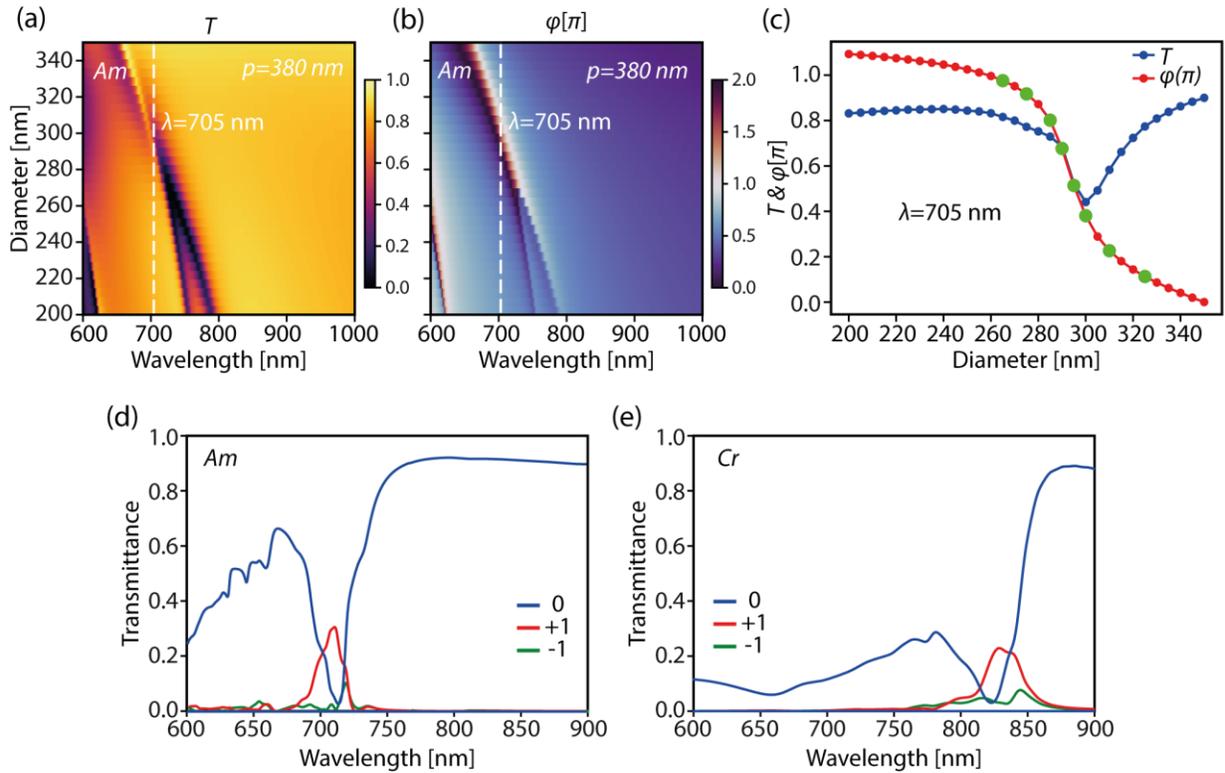

**Figure S3.** Wavefront control. a) Transmission and b) phase map of amorphous (*Am*) metasurfaces by varying nanohole diameter *d* from 200 nm to 350 nm keeping periodicity *p*=380 nm. White dashed line represents wavelength, λ=705 nm, at which the Huygens' condition is satisfied for *d* = 300 nm. c) Transmittance (*T*) and phase ($\varphi(\pi)$) plot for λ=705 nm. 8 hole diameters viz. 265, 275, 285, 290, 295, 300, 310 and 325 nm for supercell to cover complete $2\pi$ phase with incremental phase of $\pi/4$ between adjacent elements, have been highlighted by bigger green dots in the phase plot. d) Simulated transmission demonstrating beam steering in the +1 diffraction order by amorphous (*Am*) gradient metasurface with a unite cell consisting of 8 holes with different diameters, as in c), with electric field polarized in the direction perpendicular to the supercell. The beam steering efficiency reaches around 30%, at λ=710 nm, in the desired +1 order, while efficiency in the 0 and -1 order remains low. A slight shift in the λ (from 705 nm to 710 nm) pertains to the coupling between the resonances in the elements of the supercell. e) Simulated transmission of the gradient metasurface in the crystalline (*Cr*) phase. The beam

steering red-shifted to λ=829 nm with 22.8% efficiency in the +1 order, with the efficiencies in the 0 and -1 order remaining low.

**Tunable Holography**

As a final demonstration of the functionalities attainable with the tunable Huygens' metasurfaces based on nanoholes in a $Sb_2S_3$ film, we realize a switchable phase-only hologram. The 8 phase-level metasurface hologram with a footprint 300 μm × 300 μm is designed to reconstruct the image of '$Sb_2S_3$' when in the amorphous state. The phase hologram is calculated with a ray-tracing method, which is a point cloud-based computer-generated-hologram (CGH) algorithm (Figure S4a). The phase information of the hologram is digitized to eight evenly distributed phase-levels, and they are mapped to the corresponding diameter of the nanoholes in the amorphous metasurface (same as the beam steering one). A broadband supercontinuum fiber laser (SuperK EXTREME, NKT Photonics) with collimated light output incident normally on the metasurface hologram and the reconstructed image of '$Sb_2S_3$' is captured on a screen (Figure S4b). As the object is located off axis, the zeroth-order can be avoided. To characterize the efficiency of the hologram, a lens is used to collect the diffracted light and an optical power meter (Thorlabs PM320E Model Console + S120C Detector) is used to measure the power of incident and diffracted light. Note here that the measured efficiencies are purely indicative of relative efficiencies between amorphous and crystallized metasurface holograms and are given in arbitrary units as accurate efficiencies cannot be measured because the beam spot is larger than the metasurface hologram. Hence, a significant amount of light simply passes through the substrate without interacting with the metasurface, thus reducing the efficiency value. As expected, the peak efficiency is at a wavelength of 680 nm for the amorphous case. Upon crystallization, the efficiency is reduced two-and-half fold (Figure S4c). There is still some reconstruction of the hologram in the crystalline state, which we attribute to purely amplitude modulation effects, which are associated with different transmission values for holes of different sizes.

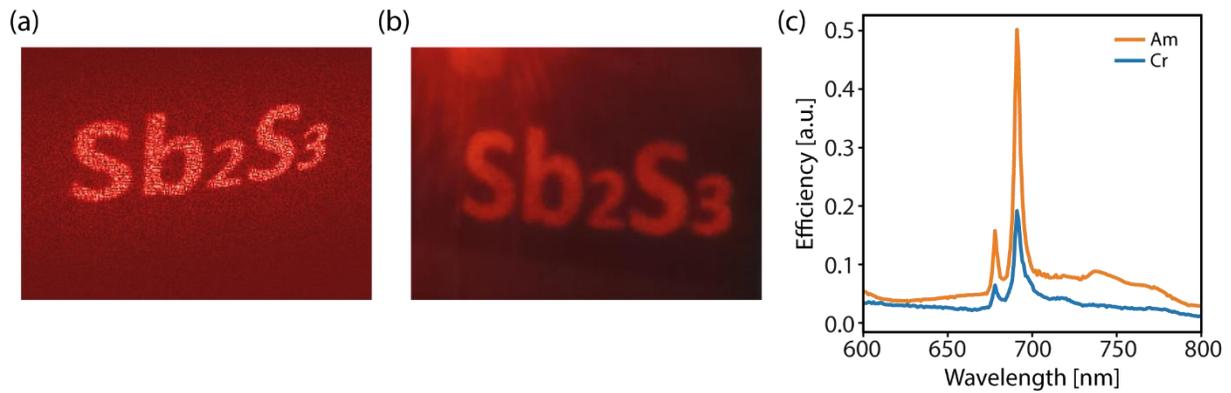

**Figure S4.** Demonstration of tunable holograms. a) calculated hologram with target reconstructed image as the symbol of antimony trisufide 'Sb$_2$S$_3$'. b) Holographic image experimentally reconstructed on a white screen distinctly showing 'Sb$_2$S$_3$' image at 680 nm in transmission through fabricated metasurface hologram in amorphous Sb$_2$S$_3$ film. c) Comparison of efficiencies between holographic images recorded for amorphous (red) and crystalline (blue) metasurface holograms.